\documentclass[pra,twocolumn,showpacs,preprintnumbers,amsmath,amssymb,superscriptaddress,unsortedaddress]{revtex4}

\pdfoutput=1
\usepackage{color}
\usepackage{graphicx}
\usepackage{dcolumn}
\usepackage{bm}

\newcommand{\al}{\alpha}
\newcommand{\be}{\beta}
\newcommand{\e}{\textrm{e}}

\newcommand{\cH}{{\mathcal H}}

\begin{document}

\preprint{}

\title{Trapping a particle of a quantum walk on the line}

\author{Antoni W\'ojcik}
\affiliation{Faculty of Physics,
 Adam Mickiewicz University,
Umultowska 85, 61-614 Pozna\'{n}, Poland.
}
\author{Tomasz {\L}uczak}
\affiliation{Faculty of Mathematics and Computer Science,
Adam Mickiewicz University,
Umultowska 87, 61-614 Pozna\'{n}, Poland.
}
\author{Pawe{\l} Kurzy\'nski}
\altaffiliation{Corresponding author.
E-mail: cqtpkk@nus.edu.sg}
\affiliation{Centre for Quantum Technologies,
National University of Singapore, 3 Science Drive 2, 117543 Singapore,
Singapore}
\affiliation{Faculty of Physics,
 Adam Mickiewicz University,
Umultowska 85, 61-614 Pozna\'{n}, Poland.
}
\author{Andrzej Grudka}%
\affiliation{Faculty of Physics,
 Adam Mickiewicz University,
Umultowska 85, 61-614 Pozna\'{n}, Poland.
}
\author{Tomasz Gdala}
\affiliation{Faculty of Mathematics and Computer Science,
 Adam Mickiewicz University,
Umultowska 87, 61-614 Pozna\'{n}, Poland.
}
\author{Ma{\l}gorzata Bednarska-Bzd\c{e}ga}
\affiliation{Faculty of Mathematics and Computer Science,
 Adam Mickiewicz University,
Umultowska 87, 61-614 Pozna\'{n}, Poland.
}

\date{\today}

\begin{abstract} 
We observe that changing a phase at a single point
in a discrete quantum walk results in a rather surprising localization
effect. For certain values of this phase change the possibility of localization strongly depends on the internal coin-state of the walker.
\end{abstract}

\maketitle

\section{Introduction}

Quantum analogs of random walks have been introduced
in hope
to device efficient algorithms of quantum computations \cite{FG} and to study new quantum effects \cite{Ahar}.
For the last decade they have
 attracted much attention  and a number of non-classical
properties of quantum walks
have been described \cite{T1,T2,T3} and experimentally verified \cite{E1,E2,E3,E4,E5,E6,E7,Broome,E9,E10}.
In particular, an enhancement and
quantum suppression typical for chaotic quantum systems
have been predicted and observed for different variants of a
one-dimensional quantum walk \cite{baker,BB,wojcik3}.

Here, we describe yet another feature of a quantum walk on an infinite chain: a
sharp allocation of a particle around a distinguished node, caused
by a local modification of the phase of a particle. We remark that
allocations in quantum walks were observed before (see e.g. \cite{L1,L2,L3,L4,L5,L6,L7,L8,L9,L10,L11}). In particular, related research include studies 
of localization in coined quantum walks with one defect coin at origin \cite{L11,Konno,CGMV}. Here we consider a simple inhomogeneity caused by a single
phase defect  at origin,
which nevertheless leads to interesting localization effects and, possibly, can be observed experimentally. We find that the number of localized stationary states strongly depends on the phase defect. We give analytical formula for these states and show that for some phase defects the effect of localization can be controlled via coin states.

Before we proceed, let us recall the standard
model of a coined quantum walk. The evolution of the system is given
by a unitary operator $U$ defined on a tensor product of two
Hilbert spaces $\cH_c \otimes \cH_n$, where $\cH_c$ is spanned by
the coin states $|c\rangle$, ($c=0,1$), while the basis of $\cH_n$
consists of vectors $|n\rangle$, ($n\in \mathbb Z$), which correspond
to the position of a particle on the chain. In a single step of the
evolution of the system we toss a coin, i.e., change the coin
state $|c\rangle$, and then move the particle to one of the two
neighboring states of the chain determined by $|c\rangle$. The corresponding unitary operator is given by
\begin{eqnarray}
U=\big(\sum_{c}|c\rangle\langle c|\otimes S_c\big)(H\otimes I),
\end{eqnarray}
where
\begin{eqnarray}
H=\frac{1}{\sqrt{2}}\left[\begin{array}{rr}1&1\\1&-1\end{array}\right]
\end{eqnarray}
is a Hadamard matrix
and $S_c$ is a translation operator defined as $S_c|n\rangle=|n+(-1)^{c+1}\rangle$.
The state vector
\begin{eqnarray}
|\Psi\rangle=\sum_n \big(\al_n |0\rangle|n\rangle+\be_n|1\rangle|n\rangle\big)
\end{eqnarray}
of the system evolves in (discrete) time $t$
according to  $|\Psi(t)\rangle=U^t|\Psi(0)\rangle$. Because the evolution of the state supported on even positions does not interfere with the evolution of the state supported on odd positions, therefore it is convenient to consider double steps, i.e., $|\Psi(t)\rangle=U^{2t}|\Psi(0)\rangle$. In this case the walk that is initially localized in position $n=0$ occupies only even positions.

\section{Single-point phase defect and localization}

Let us now break the translational symmetry of
the above model by modifying the phase of the particle by
$\omega=\omega(\phi)=\e^{2\pi i \phi}$ ($\phi\in (0,1)$) each time it passes through node $n=0$. More
precisely, we replace $U$ by $U_\phi=\big(\sum_{c}|c\rangle\langle
c|\otimes S_{c}^{\phi}\big) (C\otimes I)$, where
$S_{c}^{\phi}|n\rangle=e^{2\pi i \phi \delta_{n,0}}|n+(-1)^{c+1}\rangle$ and $\delta_{n,0}$ is Kronecker delta. We note that if the state of the coin is measured after each
step, the model corresponds precisely to the classical
random walk.

It turns out that the above symmetry breaking at node zero
results in an emergence of new localized eigenvectors of the corresponding quantum walk. These eigenvectors and the corresponding eigenvalues can be found by solving a set of recurrence equations generated by operator $U_{\phi}$. We have "atypical" equations --- these ones which involve amplitudes of probability corresponding to particle at node $0$
\begin{eqnarray}
& \sqrt{2}\alpha_{-1}(t+1)=\omega \alpha_{0}(t)+\omega \beta_{0}(t),\nonumber\\
& \sqrt{2}\beta_{1}(t+1)=\omega \alpha_{0}(t)-\omega \beta_{0}(t)
\end{eqnarray}
and "typical" equations, i.e., the ones whose right hand side does not refer to amplitudes corresponding to particle at node $0$
\begin{eqnarray}
& \sqrt{2}\alpha_{n}(t+1)=\alpha_{n+1}(t)+\beta_{n+1}(t),\nonumber\\
& \sqrt{2}\beta_{n}(t+1)=\alpha_{n-1}(t)-\beta_{n-1}(t).
\end{eqnarray}
In the above we denoted component of a state vector corresponding to particle in node $n$ and coin in state $|0\rangle$ at time $t$ by $\alpha_{n}(t)$ and component corresponding to node $n$ and coin in state $|1\rangle$ by $\beta_{n}(t)$.
For the double step operator $U_{\phi}^2$ and for stationary states these equations are time independent and take the form
\begin{eqnarray}
& 2 \lambda \bar{\alpha}_{0}=\bar{\alpha}_{1}+\bar{\beta}_{1}+\omega \bar{\alpha}_{0}-\omega \bar{\beta}_{0},\nonumber\\
& 2 \lambda \bar{\beta}_{0}=\omega \bar{\alpha}_{0}+\omega \bar{\beta}_{0}-\bar{\alpha}_{-1}+\bar{\beta}_{-1},\nonumber\\
& 2 \lambda \bar{\alpha}_{-1}=\omega \bar{\alpha}_{0}+\omega \bar{\beta}_{0}+\bar{\alpha}_{-1}-\bar{\beta}_{-1},\nonumber\\
& 2 \lambda \bar{\beta}_{1}=\bar{\alpha}_{1}+\bar{\beta}_{1}-\omega \bar{\alpha}_{0}+\omega \bar{\beta}_{0}
\label{atyp}
\end{eqnarray}
and
\begin{eqnarray}
& 2 \lambda \bar{\alpha}_{n}=\bar{\alpha}_{n+1}+\bar{\beta}_{n+1}+\bar{\alpha}_{n}-\bar{\beta}_{n} &\text{ for } n \neq 0, -1,\nonumber\\
& 2 \lambda \bar{\beta}_{n}=\bar{\alpha}_{n}+\bar{\beta}_{n}-\bar{\alpha}_{n-1}+\bar{\beta}_{n-1} &\text{ for } n \neq 0, 1,
\label{typ}
\end{eqnarray}
where $\lambda$ denotes an eigenvalue of $U_{\phi}^2$ and for convenience we substituted  $\bar{\alpha}_{n}=\alpha_{2n}$ and $\bar{\beta}_{n}=\beta_{2n}$.

The solution to Eqs. \ref{atyp} and \ref{typ} (for detailed derivation see the Appendix) yields

\begin{eqnarray}
\lambda_{\pm}=\frac{\omega-2\omega^{2}+\omega^{3}\pm i \omega(1-\omega+\omega^{2})}{1-2\omega+2\omega^{2}},
\end{eqnarray}
whereas the components of the localized stationary state vectors which do not correspond to the particle at node $0$ are given by
\begin{eqnarray}
\bar{\alpha}_{n}^{(\pm)} &= C x_{\pm}^{n}~~~~~~~~~~~~~~~~~ & \text{ for } n \geq 1,\nonumber\\
\bar{\alpha}_{n}^{(\pm)} &= C (\omega\mp i \omega \pm i) x_{\pm}^{-n} &\text{ for } n \leq -1 \label{alpha}
\end{eqnarray}
and
\begin{eqnarray}
\bar{\beta}_{n}^{(\pm)} &= C(1-\omega\mp i\omega)x_{\pm}^{n} &\text{ for } n \geq 1,\nonumber\\
\bar{\beta}_{n}^{(\pm)} &= \mp i C x_{\pm}^{-n}~~~~~~~~~~~~ &\text{ for } n \leq -1, \label{beta}
\end{eqnarray}
where
\begin{eqnarray}\label{x}
x_{\pm}=\frac{1}{2 \cos(2 \pi \phi) \mp 2 \sin(2 \pi \phi) - 3}.
\end{eqnarray}
The components of the localized stationary state vectors which correspond to the particle at node $0$ are
\begin{eqnarray}
\bar{\alpha}_{0}^{(\pm)} & = & C,\nonumber \\
\bar{\beta}_{0}^{(\pm)} & = & \mp i C. \label{zero}
\end{eqnarray}
In the above $C$ is a normalization constant 
\begin{equation}\label{norm}
C=\sqrt{\frac{1+x_{\pm}}{2}}. 
\end{equation}
It follows that there are at most two stationary states to which we refer as to $|\Phi_{\pm}\rangle$.

\section{Discussion}

\begin{figure}
\scalebox{1.0}
{\includegraphics{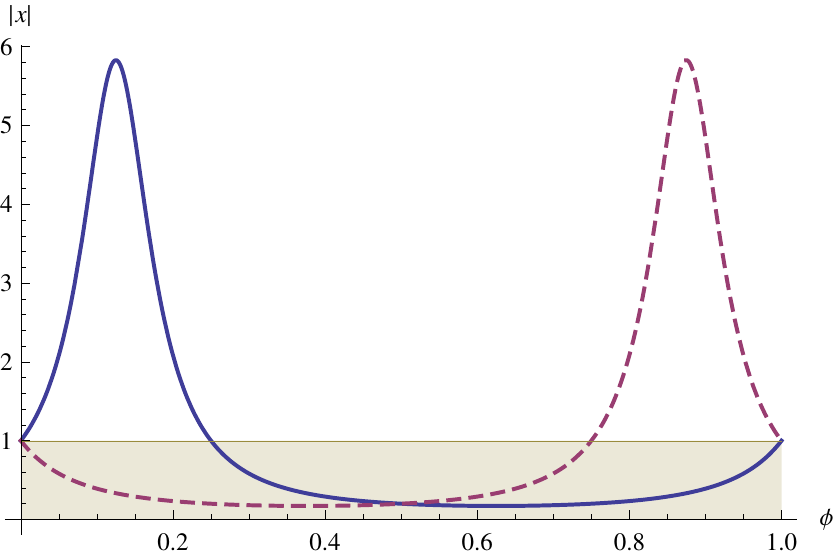}}
\caption{\label{f1} The plots of absolute values of $x_+$ (dashed) and $x_-$ (solid) as a function of $\phi$. The shaded region denotes the allowed values of $x_{\pm}$. $|x_-|<1$ for $\phi\in \left(\frac{1}{4},1\right)$ whereas  $|x_+|<1$ for $\phi\in \left(0,\frac{3}{4}\right)$.}
\end{figure}
 
First, let us observe that the number of stationary states depends on $\phi$. Due to the normalization condition both $\bar{\alpha}_n$ and $\bar{\beta}_n$ should tend to zero as $n$ tends to $\pm \infty$. Therefore, the parameter $x_{\pm}$ defined in (\ref{x}) has to obey $|x_{\pm}|<1$. This is true only for certain values of $\phi$ (see Fig. \ref{f1}), namely $|x_-|<1$ for $\phi\in \left(\frac{1}{4},1\right)$ whereas  $|x_+|<1$ for $\phi\in \left(0,\frac{3}{4}\right)$. Therefore, in the range $\phi\in (0,\frac{1}{4}]$ there is only one localized stationary state $|\Phi_+\rangle$, in the range $\phi\in(\frac{1}{4},\frac{3}{4})$ there are two such states $|\Phi_+\rangle$ and $|\Phi_-\rangle$, and in the range $\phi\in [\frac{3}{4},1)$ there is again one localized stationary state, but this time it is $|\Phi_-\rangle$. 

Next, let us study the localization properties of the quantum walk with single-point phase defect. We say that the walk exhibits localization if in the limit of infinite (even) time the probability of occupying the initial position does not tend to zero \cite{Konno}. From the previous paragraph we know that there are at most two normalizable eigenstates of  $U_{\phi}^2$ that correspond to stationary states $|\Phi_{\pm}\rangle$. By the analogy to textbook problems of a quantum particle in the presence of a potential trap, these states can be thought of as bound states. The remaining states correspond to scattering states. We expect that an initially localized state whose decomposition into eigenstates of the evolution operator does not contain bound states does not exhibit localization (see \cite{Konno} and references therein). Therefore, the overlap of the initial state of the walk with bound eigenstates can be used to measure the localization properties of the walk. Note that this overlap, which we denote as $F$, is definitely a lower bound for the probability of occupation of the initial position in the limit of infinite time.

\begin{figure}
\scalebox{1.0}
{\includegraphics{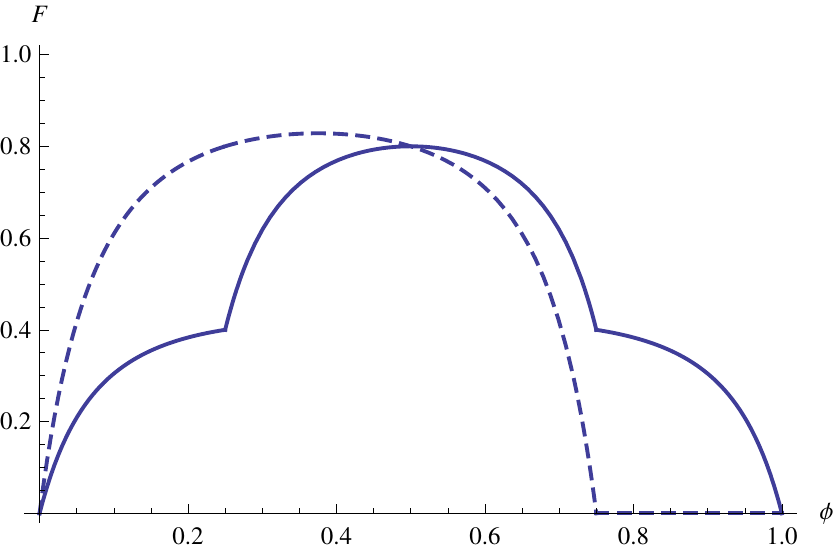}}
\caption{\label{f2} The plots of overlap $F$ of initial states of the quantum walk and the localized stationary states of $U_{\phi}^2$. The solid line corresponds to the initial coin state $|0\rangle$, whereas the dashed line corresponds to the initial coin state  $\frac{1}{\sqrt{2}}(|0\rangle+i|1\rangle)$.}
\end{figure}

For initial state of the form $|\psi_{0}\rangle=\alpha|0\rangle|n=0\rangle+\beta|1\rangle|n=0\rangle$,  the total overlap is:
\begin{eqnarray}
& F=F_+ \text{ for } \phi\in (0,\frac{1}{4}]\nonumber\\
& F=F_+ + F_- \text{ for } \phi\in(\frac{1}{4},\frac{3}{4})\nonumber\\
& F=F_- \text{ for } \phi\in [\frac{3}{4},1),
\end{eqnarray}
where
\begin{equation}\label{overlap}
F_{\pm}=|\langle\Phi_{\pm}|\psi_0\rangle|^2=\frac{1+x_{\pm}}{2}|\alpha \pm i\beta|^2.
\end{equation}
Fig. \ref{f2} shows the plots of $F$ for initial coin states $|0\rangle$ and $\frac{1}{\sqrt{2}}(|0\rangle +i|1\rangle)$.
This example evidently shows that localization properties depend on the initial coin state of the particle. In particular, there are two initial states  $|\psi_{0}^{\pm}\rangle=\frac{1}{\sqrt{2}}(|0\rangle|n=0\rangle\pm i|1\rangle|n=0\rangle)$ that are orthogonal to localized stationary states $|\Phi_{\mp}\rangle$, respectively. As a result, we observe that the initial state $|\psi_0^{+}\rangle$ exhibits localization only in the range $\phi\in \left(0,\frac{3}{4}\right)$, whereas the state $|\psi_0^{-}\rangle$ exhibits localization only in the range $\phi\in \left(\frac{1}{4},1\right)$. 
\begin{figure}
\scalebox{1.0}
{\includegraphics{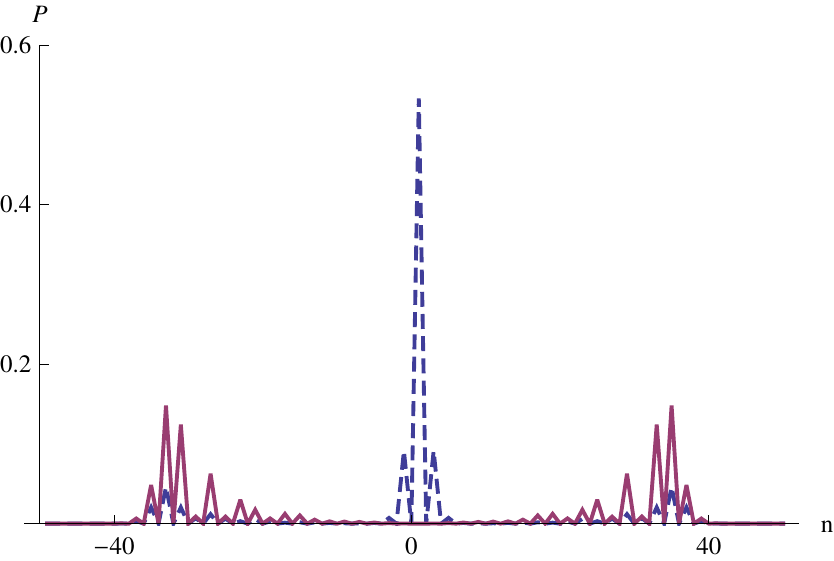}}
\caption{\label{f3} Probability distribution after 50 steps of a quantum walk with a single phase defect at origin corresponding to $\phi=\frac{1}{6}$. Solid line --- lack of localization for the initial coin state $\frac{1}{\sqrt{2}}(|0\rangle-i|1\rangle)$. Dashed line ---  localization at origin for the initial coin state $\frac{1}{\sqrt{2}}(|0\rangle+i|1\rangle)$.}
\end{figure} 
In the Fig. \ref{f3} we present probability distributions for a quantum walk with phase defect at origin corresponding to $\phi=\frac{1}{6}$ for initial states $|\psi_0^{+}\rangle$ for which both localization and the lack of localization are well visible.

\section{Experimental proposal}

Here we propose an experiment to verify our theoretical results. It is a simple modification of the experiment performed by Broome {\it et. al.} \cite{Broome} that employs a single photon "walking" in an array of beam splitters (beam displacers) where the polarization of the photon plays the role of the coin. We believe that our proposal can be relatively easy accommodated to any other quantum walk implementation. 
\begin{figure}
\scalebox{0.4}
{\includegraphics{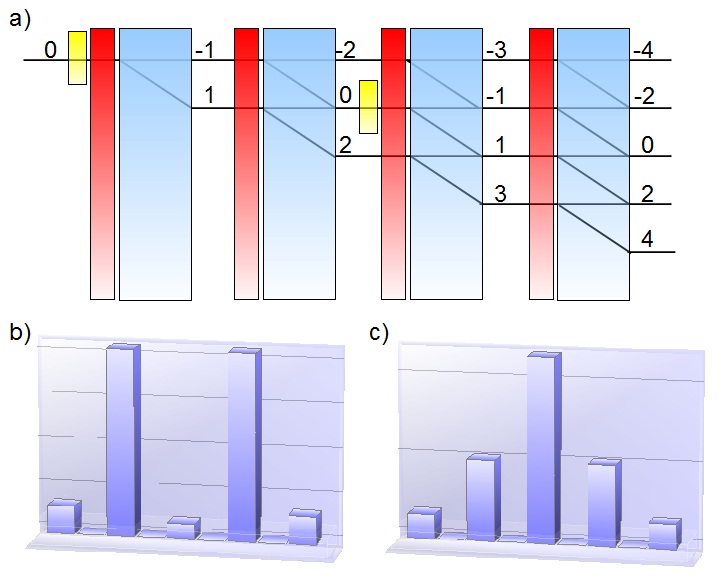}}
\caption{\label{f4} a) Experimental proposal for four steps of a quantum walk with a phase defect at position $n=0$. It employs a single photon in an array of optical elements and is based on a setup introduced in \cite{Broome}. Bold (blue) extended rectangles represent calcite beam displacers which realize a coin-dependent position shift, thin (red) extended rectangles represent half-wave plates/quarter-wave plates which realize coin operations, and small (yellow) rectangles represent phase shifters which implement a phase shift at position $n=0$. Below we present expected probability distributions after four steps for a phase shift corresponding to $\phi=\frac{1}{6}$. b) Initial coin state $\frac{1}{\sqrt{2}}(|0\rangle-i|1\rangle)$. c) Initial coin state $\frac{1}{\sqrt{2}}(|0\rangle+i|1\rangle)$.}
\end{figure} 
A pair of photons is created in a process of spontaneous parametric down-conversion. The presence of the photon that is going to be injected into an array of calcite beam displacers and wave plates is heralded by a detection event of its twin. Before the walk starts, the initial state of the polarization (coin) is prepared via application of a proper wave plate. The scheme of an array implementing a quantum walk with a single phase defect at position $n=0$ is presented in Fig. \ref{f4}. Calcite crystals displace the path of the photon with respect to its polarization state, therefore they are capable of realizing a coin-dependent position shift. Wave plates change the polarization state, hence they realize the coin operation. Finally, properly placed phase shifters can implement a phase shift at origin.

Due to the fact that in realistic experiments one always encounters inevitable errors it is of great importance to determine the conditions under which the investigated phenomena is observable. It is obvious that the more steps of a quantum walk one wants to implement, the more robust to noise the experimental setup has to be. Fortunately for us, the first signs of the coin-dependent localization can be observed after only four steps of the corresponding walk (see Fig. \ref{f4} b and c). Note that the authors of \cite{Broome} implemented six steps.

An interesting addition to the experiment would be to investigate the effects of decoherence on localization. It is well known that decoherence transforms quantum walks into classical random walks \cite{decoh} for which interference effects play no role, hence the phase shift at position $n=0$ would cause no effect and the corresponding walk should exhibit Gaussian pattern irrespective of the initial coin state. Note that tunable decehorence is within experimental reach since the authors of \cite{Broome} implemented it via change of relative angles between calcite crystals.

\section{Further research}

The state-dependence of localization can be used to filter and to trap particles of discrete time quantum walks with respect to their internal coin state. This effect can find applications in quantum algorithms and in quantum state engineering.  Another interesting perspective is to investigate localization due to a single phase defect in quantum walks in two (and more) dimensions. In particular, it would be interesting to investigate single-point localization on grids and honeycomb lattices, since the former represent natural generalization of a one-dimensional chain and the latter are interesting from the point of view of the recent intensive research on the material properties of graphene.

\section{Acknowledgements}

T{\L} was partially supported by the Foundation for Polish Science. PK was supported by the National Research Foundation and Ministry of Education in Singapore.

\subsection{Appendix}

Here we give detailed solution to Eqs. (\ref{atyp}) and (\ref{typ}). Recall that we use the following substitution $\bar{\alpha}_{n}=\alpha_{2n}$ and $\bar{\beta}_{n}=\beta_{2n}$. From Eqs. (\ref{typ}) one obtains
\begin{eqnarray}\label{rec}
\bar{\beta}_{n+1}=\frac{1}{\lambda-1}(\bar{\alpha}_{n+1}-\lambda \bar{\alpha}_{n}) \text{ for } n\neq-1,0
\end{eqnarray}
Substituting the above into the second equation in (\ref{typ}) we have
\begin{eqnarray}
& \lambda \bar{\alpha}_{n+1}-2(\lambda^{2}-\lambda+1)\bar{\alpha}_{n}+\lambda\bar{\alpha}_{n-1}=0\nonumber\\
& \text{ for } n\neq -1,0,1
\end{eqnarray}
The general solution to this recurrence equation is of the form
\begin{eqnarray}
\bar{\alpha}_{n}=C_{+}x^{n}+C_{-}x^{-n}, \nonumber
\end{eqnarray}
where $x$ as well as $x^{-1}$ satisfy equation
\begin{eqnarray} \label{ceq}
\lambda x^{2}-2(\lambda^{2}-\lambda+1)x+\lambda=0 \text{ for } i\neq-1,0,1
\end{eqnarray}
and $C_{+}$, $C_{-}$ are constants. Due to normalization constraint for $n \rightarrow \pm \infty$ we must have $|\bar{\alpha}_{n}| \rightarrow 0$, therefore
\begin{eqnarray}
& \bar{\alpha}_{n}=C_{+}x^{n} \text{ for } n \geq 1,\nonumber\\
& \bar{\alpha}_{n}=C_{-}x^{-n} \text{ for } n \leq -1, \label{alp}
\end{eqnarray}
where $x$ is the solution to (\ref{ceq}) that satisfies $|x|<1$.
Substituting Eqs. (\ref{alp}) into Eq. (\ref{rec}) we have
\begin{eqnarray}
& \bar{\beta}_{n}=C_{+}\frac{x-\lambda}{\lambda-1}x^{n-1} \text{ for } n \geq 2,\nonumber\\
& \bar{\beta}_{n}=C_{-}\frac{1-\lambda x}{\lambda-1}x^{-n} \text{ for } n \leq -1. \label{bet}
\end{eqnarray}
From the first and the fourth equation of (\ref{atyp}) we obtain
\begin{eqnarray}
\bar{\alpha}_{0}=\frac{1}{\lambda}(\bar{\alpha}_{1}+\bar{\beta}_{1}(1-\lambda)).
\end{eqnarray}
Substituting Eqs. (\ref{alp}) and (\ref{bet}) into the above we have
\begin{eqnarray}
\bar{\alpha}_{0}=C_{+}.
\end{eqnarray}
Observe that due to the above the first equation in (\ref{bet}) is also valid for $n=1$. From the second and the third equation of (\ref{atyp}) we obtain
\begin{eqnarray}
\bar{\beta}_{0}=\frac{1}{\lambda}(\bar{\beta}_{-1}+\bar{\alpha}_{-1}(\lambda-1)).
\end{eqnarray}
From  Eqs. (\ref{alp}), (\ref{bet}) and (\ref{ceq}) it follows that
\begin{eqnarray}\label{b0}
\bar{\beta}_{0}=C_{-}\frac{1-\lambda x}{\lambda-1}.
\end{eqnarray}
From Eqs. (\ref{alp}) and (\ref{bet}) and from the first and the second equation of (\ref{atyp}) we have
\begin{eqnarray}
& \omega \bar{\beta}_0=C_{+}(x+\omega-2\lambda+\frac{x-\lambda}{\lambda-1}),\nonumber\\
& C_{+}\omega=\bar{\beta}_0(2\lambda-x-\omega+x\frac{\lambda-1}{1-\lambda x}), \label{C+}
\end{eqnarray}
which together give
\begin{eqnarray}
\omega^{2}=(x+\omega-2\lambda+\frac{x-\lambda}{\lambda-1})(2\lambda-x-\omega+x\frac{\lambda-1}{1-\lambda x}).
\end{eqnarray}
Taking into account Eq. (\ref{ceq}) the solution to the above equation yields
\begin{eqnarray}\label{sol}
\lambda_{\pm}=\frac{\omega-2\omega^{2}+\omega^{3}\pm i \omega(1-\omega+\omega^{2})}{1-2\omega+2\omega^{2}}
\end{eqnarray}
and
\begin{eqnarray}
x_{\pm}=\frac{\omega}{\omega^{2}-3\omega+1\pm i(\omega^{2}-1)},
\end{eqnarray}
which can be expressed as
\begin{eqnarray}\label{sol2}
x_{\pm}=\frac{1}{2 \cos(2 \pi \phi) \mp 2 \sin(2 \pi \phi) - 3}.
\end{eqnarray}
From Eqs. (\ref{b0}), (\ref{C+}), (\ref{sol}) and (\ref{sol2}) one finds
\begin{equation}
C_-=C_+(\omega \mp i\omega \pm i),
\end{equation}
where the $\pm$ sign depends on the signs in Eqs. (\ref{sol}) and (\ref{sol2}). As a consequence, we arrive at 
\begin{eqnarray}
\bar{\alpha}_{n}^{(\pm)} &= C x_{\pm}^{n}~~~~~~~~~~~~~~~~~ & \text{ for } n \geq 1,\nonumber\\
\bar{\alpha}_{n}^{(\pm)} &= C (\omega\mp i \omega \pm i) x_{\pm}^{-n} &\text{ for } n \leq -1, \nonumber \\
\bar{\beta}_{n}^{(\pm)} &= C(1-\omega\mp i\omega)x_{\pm}^{n} &\text{ for } n \geq 1,\nonumber\\
\bar{\beta}_{n}^{(\pm)} &= \mp i C x_{\pm}^{-n}~~~~~~~~~~~~ &\text{ for } n \leq -1
\end{eqnarray}
and
\begin{eqnarray}
\bar{\alpha}_{0}^{(\pm)} & = & C,\nonumber \\
\bar{\beta}_{0}^{(\pm)} & = & \mp i C,
\end{eqnarray}
where we substituted $C=C_+$. The normalization constant $C$ can be easily evaluated since the sum of probabilities $|\bar{\alpha}_n|^2$ and $|\bar{\beta}_n|^2$ gives rise to the sum of geometric series. It yields
\begin{equation}
C=\sqrt{\frac{1+x_{\pm}}{2}}.
\end{equation}

\end{document}